# Packing and voids in electro-rheological structures of polarized clay particles

K. P. S. Parmar [a)], Y. Méheust, and J. O. Fossum[a)]
*Physics Department, Norwegian University of Science and Technology (NTNU), Trondheim, Norway*

K. D. Knudsen
*Physics Department, Institute for Energy Technology (IFE), Kjeller (Norway)*

D. M. Fonseca
*Physics Department, Norwegian University of Science and Technology (NTNU), Trondheim, Norway*

Oil suspensions of fluorohectorite clay particles exhibit a dramatic meso-structural ordering when submitted to a strong electric field. This is due to dipolar interaction between polarized fluorohectorite particles, which orientate and aggregate to forms chains and/or bundle-like structures along the direction of applied electric field. We have used synchrotron small angle X-ray scattering to get insight into the nature of the porous medium in the bundles. Three types of fluorohectorite clay samples corresponding to three different intercalated cations $Na^+$, $Ni^{2+}$ and $Fe^{3+}$ were studied. The two-dimensional SAXS images from bundles of fluorohectorites exhibit a marked anisotropy which is analyzed by fitting ellipses to iso-intensity lines of SAXS patterns. This also provides principal directions along which one-dimensional spectra are computed. They display a power law behavior typical of porous media, separated by crossovers. The crossovers are interpreted in terms of typical length scales for the clay particle bundles, providing for the first time a quantitative image of the 3D geometry inside such bundles of polarized clay particles. The exponents of the power laws indicate either predominant surface- (for 2 types of samples) or bulk- (for the last type) scattering, at all length scales investigated.

---

[a)] Emails: kanak.parmar@phys.ntnu.no, jon.fossum@ntnu.no

## I. INTRODUCTION

The rheological properties (viscosity, yield stress, shear modulus etc.) of colloidal suspensions consisting of small particles of dielectric constant and/ or conductivity[1–3] higher than that of the surrounding liquid can be controlled by an external electric field. Such suspensions are known as electrorheological (ER) fluids[4, 5]. The applied electric field (of the order of kV/mm) results in induced dipoles in the particles. Consequently, the particles undergo positional and orientational alignment to minimize dipole interaction energy, which results in their aggregation into chain/bundle-like structures, parallel to the direction of applied electric field[6–10].

The study of the structure formation of ER fluids has attracted increasing interest in recent years for its fundamental and technological importance. It has been shown that microcrystalline structures inside the bundles of spherical particles are a body-centered tetragonal (bct) lattice[11–13]. This provides a new technique to form mesocrystals with unique photonic properties[14-15]. The details of the structure formation in ER fluids depend on several factors, such as concentrations[16], size distributions of particles[17], field frequency[18-19] etc. The influence of the dielectric constant or conductivity of particles is another interesting topic since the dielectric and conductivity mismatch between the particles and the liquid is widely accepted as the main cause of the ER phenomenon. The yield stress of an ER fluid is a critical parameter that governs the numerous potential applications of such a fluid. Besides the other experimental parameters, it depends upon the internal close packing of the particles density[20].

In this paper, we address ER fluids consisting of 2:1 clay particles suspended in silicone oil. Among the large variety of clay minerals, 2:1 clays have been investigated for potential applications such as nanocomposites, catalysts, cosmetics and pharmaceuticals etc., and applications where their unique bonding, suspending or gellant properties are required. The basic physical structural unit of a 2:1 clay is a platelet (about 1 nm thick) made of one aluminum or magnesium hydroxide octahedral sheet sandwiched between two inverted tetrahedral silica sheets. The isomorphic substitutions in the octahedral and or tetrahedral sheets by less charged cations generate a net negative layer charge on the platelet[21]. To balance this structural layer charge, individual platelets tend to stack by sharing intercalated (possibly hydrated) cations. The structural charge on the platelets of 2:1 varies from 0.4 to 1.2 e- per unit cell[22]. This moderate platelet charge results in physical and chemical properties that are unique and not found in other clay minerals.

Fluorohectorite is a synthetic 2:1 clay that has one of the largest values for the particle diameter, up to about 10 μm, and a layer charge of 1.2 e$^-$ per unit cell, which is very large in comparison to that of other synthetic 2:1 such as laponite (0.4 e$^-$ per unit cell)[23]. This high surface charge leads to a very robust stacking of the fluorohectorite platelets, resulting in fluorohectorite particles effectively consisting of ~ 100 stacked platelets[24-25]. Those particles (or grains) are ~0.1 μm-thick; they remain quasi-two-dimensional in nature, as the lateral dimensions may be as much as 200 times the particle thickness. However, fluorohectorite is polydisperse with a wide distribution in lateral sizes, so that this ratio is not fixed, and may be considerably lower for other grains.

Colloidal suspensions of fluorohectorite particles dispersed in a low conductivity silicone oil have been shown to exhibit a dramatic electro-rheological behavior[26]. A wide-angle X- ray scattering (WAXS) study of the ER bundles has demonstrated the clay particles' preferred orientations inside the bundles, and shown that they polarize along their silica sheets[26]. The particular way the anisotropic clay particles organize in the bundle with respect to one another i.e., their positional and orientational order, determines the morphology of the porous medium in-between the aggregated particles.

Small angle X-ray scattering (SAXS) and small angle neutron scattering (SANS) are techniques commonly used to study the mesoscopic structural arrangement of soft matter. In particular, they have proved very useful in the study of systems of clay mineral particles in aqueous media, for which a variety of structures/properties under different preparation conditions have been demonstrated, such as spontaneous ordering, aggregation, gelation and thixotropy behavior[27–30]. In this paper we have used SAXS to study the electrorheological bundles of clay particles in silicone oil in order to get insight into the nature of the porous medium in between the clay particles that make up the chain bundles.

Following this introduction, the paper is structured as follows: section 2 addresses the features of small-angle scattering spectra expected from a meso-porous medium; section 3 contains a description of the experiments and a discussion of their results; we conclude in section 4.

## II. SMALL-ANGLE SCATTERING IN A POROUS MEDIUM

The fluorohectorite particles consist of stacks of 1 nm- thick platelets with a thickness ~100 nm and a lateral extension between 200 nm and 2 μm. X-ray diffraction from the stacked structure is classically used to determine the platelet separation of such nano-stacks[31-32], and their thickness. More recently, it has been used to infer particle orientation distributions in poly-crystalline samples of these nano-stacks[33]. Besides wide angle diffraction (WAXS), the dimensions of the clay particles make small angle scattering a suitable tool when it comes to studying the collective behavior/organization of these clay particles.

Small-angle scattering is a useful tool to study disordered structures and porous media[34–43]. In a small angle X-ray scattering (SAXS) experiment, the dependence of the scattered intensity on the scattering angle $2\theta$ is controlled by the size of the colloidal particles, their tendency to aggregate, the porosity of the disperse system, the magnitude of the specific surface area, and more generally, by the inhomogeneties characterizing the structure of the disperse system[44-45]. In the Born approximation, the scattered intensity $I(q)$, as a function of the momentum transfer vector $q$

$$q = \frac{4\pi}{\lambda} \sin\left(\frac{2\theta}{2}\right) \tag{0}$$

where $\lambda$ is the wavelength of the incident beam and $2\theta$ is the scattering angle, is simply proportional to the Fourier transform of the geometric correlation function of the electron density.

For porous media, the scattering profiles often exhibit power laws on a wide $q$-range. This has been observed in particular for aggregates of colloids[36–38], carbons[46] and various types of rocks[44-45, 47–49]. Such power law scattering curves have been explained in terms of fractal properties of the diffracting medium, from models in which aggregated particles (for example, ruguous spheres) form a porous medium with geometrical properties that relate to the concept of fractals. Here, it is important to distinguish between mass and surface fractals[50].

If we consider mass fractals, or more generally systems in which the aggregated particles are positioned with respect to one another so that their two-point correlation function is a power law function in the form

$$g(R) \propto R^{D_m-3} \qquad (2)$$

where $D_m$ is a fractal dimension, a power law scattering curve with an exponent $-D_m$ is observed for $q$-values in the range $1/l \leq q \leq 1/a$, if $a$ is the typical size of an aggregated particle, and $l$ is the size of the whole system, i.e., the largest accessible length scale[43, 50]. Since by definition $D_m$ is in the range [1; 3], the exponent observed in the scattering data is in the range [−3; −1].

If we now probe length scales smaller than the particle size, the scattering becomes sensitive to the roughness of the aggregated particles. If this roughness qualifies the particle for being considered a surface fractal with a dimension $D_s$, or more generally if the area is a power law of the horizontal size with an exponent $D_s$ on a given range of scales, then the scattering profiles exhibit a power law behavior with an exponent $-(6-D_s)$ on that range of scales[44, 50]. Here by definition $D_s$ is in the range [2; 3], and the scattering curve exponent is in the range [−4; −3], with the well-known Porod law limit −4 being reached for smooth particles[42].

In conclusion, in such systems of aggregated rough particles, two power law behaviors are observed, one with an exponent between −4 and −3 at large $q$ - values, and one with an exponent between −3 and −2 at intermediate $q$ - values corresponding to length scales larger than the particle size. The porous medium consisting of aggregated clay particles that we address in the present study certainly bears similarities with the models presented above, and will be studied in this framework. The scattering properties of porous media possessing long range correlations in their mass geometry or in the roughness of their pores are generally investigated in this framework as well, and fitted a power law model in the form

$$I(q) = Aq^{-D} + B \qquad (3)$$

where $A$ is a scaling factor and $B$ is the incoherent background scattering. However, the understanding of such data is more difficult in the presence of a polydispersity of the aggregated particles, or, equivalently, of a wide distribution for the pore sizes. In this case, an exponent (in absolute value) $D > 3$ can arise both from surface scattering form rough pore walls, or from bulk scattering from a porous media with a wide pore size distribution[51].

## III. EXPERIMENTS

### III.1. EXPERIMENTAL METHOD

Synthetic sodium fluorohectorite, with a chemical formula $Na_{0.6}(Mg_{2.6} Li_{0.6}) Si_4 O_{10} F_2$ per unit cell (where Na is an interlayer exchangeable cation) was purchased form Corning Inc. (New York) in powder form. Three fluorohectorite clay powders, with different intercalated cations, sodium ($Na^+$), nickel ($Ni^{2+}$) and iron ($Fe^{3+}$), were prepared through an ion exchange method: the corresponding cation salts were dissolved in distilled water and added in an amount approximately five times the cation exchange capacity in order to promote saturation. After thorough mixing, three fluorohectorite clays were dialyzed against distilled water until a negative chloride test was obtained. After dialysis, the three fluorohectorite clays were dried out and grinded in powder form. Suspensions at 3 % (w/w) were then prepared by directly mixing the chosen amount of clay powder with Rothiterm silicone oil in sealed glass tubes at ambient temperature. Finally, the mixtures were shaken vigorously for 15 min, and subsequently left to rest.

The synchrotron SAXS experiments were performed using the supernatant of these suspensions in a custom-made scattering cell. The scattering cell was made of an insulating plastic material, whose top part was open and both the (front and back) sides and the bottom part were closed by gluing a standard thin (thickness, 0.08mm) kapton film. Two parallel and identical 1/2 mm-thick copper electrodes separated by a uniform gap of 2 mm were inserted from the top of the sample cell. The suspension to be studied (< 2 ml) was introduced between the electrodes from the top part. After application of an electric potential difference (2 kV) between these copper electrodes, the clay particles were first observed to form chain-like structures parallel to the applied electric field. The chains were subsequently observed to collapse, forming stable chains bundles in about 10 sec. Fig. 1 shows one such structure formed by iron ($Fe^{3+}$) - fluorohectorite clay particles.

SAXS experiments were carried out under ambient temperature at beamline BM-26B of the European Synchrotron Radiation Facility (ESRF), in Grenoble (France), using a wavelength $\lambda$ = 1.24 Å. Scattered X- rays were recorded on a two-dimensional detector (a square matrix of 512×512 elements) positioned about 8000 mm away from the sample; the exposure time was 600 s. The experimental geometry is shown in Fig. 2: the ER bundles were lying on average horizontally. In what follows, we shall refer to the corresponding axis of the SAXS images as the horizontal, while the perpendicular one shall be denoted as the vertical direction. This experimental setup allowed SAXS data on a range of momentum transfer $q$-values between $q_{min}$ = 0.01 nm$^{-1}$ up to $q_{max}$ = 0.15 nm$^{-1}$, corresponding to length scales $d_{max}$ ~ 600 nm and $d_{min}$ ~ 40 nm, respectively.

## III.2. DATA ANALYSIS

### III.2.1. METHOD

The raw two-dimensional SAXS patterns obtained from bundle structures for all the three samples – sodium ($Na^+$) fluorohectorite, nickel ($Ni^{2+}$) fluorohectorite and iron ($Fe^{3+}$) fluorohectorite are clearly anisotropic (see Fig. 3). The iso-intensity lines are roughly homothetic and resemble ellipses that would have their major axis lying approximately vertical, and their minor axis lying approximately horizontal. We first analyzed the two-dimensional SAXS images using a home-made image analysis software in Matlab. Ellipses were fitted to iso-intensity lines for each SAXS pattern; the principal axes (direction and length) were computed, and an eccentricity $e$ defined as

$$e = \sqrt{1 - \frac{j^2}{k^2}} \quad , \tag{4}$$

where $j$ and $k$ are the long and short axes of the fitted ellipse, was determined. This eccentricity accounts for the anisotropy of the images. The eccentricity values obtained for each type of sample under an identical electric field are shown in Table 1. Those values are average values obtained from several images recorded from identical samples; this allowed us to account for the large variations arising from the noise in the data. For each eccentricity value in Table 1, the number of images used to calculate it, is also listed.

For further analysis, each SAXS pattern was corrected for detector sensitivity and background by subtracting from the image data an image data of the empty cell.

Besides, the sample to detector distance was calibrated using the fiber diffraction of wet rat-tail collagen, which has a strong characteristic peaks at $q = 2\pi n/67.2$ nm$^{-1}$ ($n = 1, 3, 5$). We then computed one-dimensional profiles of X-ray intensity $I(q)$ versus scattering vector $q$ along the principal axes of the ellipses, as obtained previously. Those directions are nearly parallel to the horizontal (i.e. the direction of the applied electric field) and the vertical directions, respectively. The profiles were obtained by averaging the two-dimensional images over azimuthal angles, in ± 5°- wide sectors around those two directions. Log-log plots of those profiles appear to consist of linear segments, which mean that different power laws in the form of Eq. (3) are observed on different $q$-ranges. Fig. 4 (a) and Fig. 4 (b) illustrate this behavior for profiles recorded along the minor- and the major- axis respectively.

For intensity profiles recorded along the horizontal direction (Fig. 4 (a)) three different $q$-ranges are clearly visible, on which three power laws can be fitted. The crossover $q$- values were first estimated visually, with an uncertainty; they are listed as $q_{\parallel i}$ (i = 1, 2) in Table 1, the index $i$ being incremented as one moves from larger to smaller $q$-values. Straight lines were fitted to the linear segments in the log-log plots; they are shown Fig. 4 (a) while the corresponding power law exponents $D_{\parallel i}$ (i = 1, 2, 3) (same convention on the index $i$ as for the $q_i$s) are listed together in Table 1.

For intensity profiles recorded along the vertical direction (see Fig. 4(b)), we also distinguish three regimes, but the crossovers are somewhat more difficult to determine than along the horizontal axis. The crossovers $q_{\perp 1}$ are determined visually, and the crossovers $q_{\perp 2}$ are chosen equal to $q_{\parallel 2}$ (see section 3.2.2). The fitted values of the exponents $D$ differ enough from each other on the adjacent ranges (see the $D_{\perp i}$, $i = 1, 2$ in Table 1) to confirm the existence of two crossovers. As explained above for the ellipse eccentricity values, the values for the exponents $D$ listed in Table 1 are obtained from averaging over 7 images recorded on identical samples. Note that no power law exponent was fitted on the low-q ranges in Fig.4(b), because the data range was not wide enough, and did not allow good enough confidence levels on the fits.

**III.2.2. RESULTS AND DISCUSSIONS**

Table 1 lists all results, both related to the eccentricities of the two-dimensional images and to power law exponents of one-dimensional scattering profiles.

The minor axis of the ellipses fitted to the two-dimensional SAXS images are close to horizontal. This shows that the apparent size of the clay particles along that direction, as probed by the X-ray, is larger along that horizontal direction – also the direction of the electric field – than along the vertical direction. Since the clay particles are platelets with a thickness much smaller than their lateral dimension, this suggests that they are lying in the ER bundles with one of their lateral dimensions parallel to the field; i.e., on average, the clay particles have their directors aligned perpendicular to the direction of applied field. The maximum eccentricity is seen for nickel (Ni) fluorohectorite ($e = 0.61$), and the minimum eccentricity for sodium (Na) fluorohectorite ($e = 0.54$). The difference is less than 12% (see Table 1). This means that, at equal electric field strength, the orientational order for the particles are of the same order of magnitude for all types of samples.

If all the particle directors were aligned with the vertical direction, the anisotropy of the SAXS pictures would be much larger than what is observed. Whether the directors are aligned with each other along another direction perpendicular to that of the applied field, or whether they are widely distributed (and possibly, uniformly) in the vertical plane that contains the incoming beam, cannot be inferred directly from the SAXS data. In a recent study, however, we investigated the same electro-rheological bundles using X-ray diffraction (i.e., wide angle X-ray scattering, or WAXS)[26]. Taking advantage of the nano-layered nature of the clay crystallites, we were able to infer particle orientations inside the particle bundles from the anisotropy of two-dimensional WAXS images[33]. This study suggested that the particles polarize along their silica sheets (see Fig. 1(c)), and hence, that their directors were, on average, perpendicular to the direction of the electric field. Our SAXS data is consistent with these previous findings (see second paragraph of section 3.2.2). In Figure 1(b), we present a two-dimensional sketch of an aggregate inside the particle assembly, as inferred from the study in[26] and from the current SAXS study. In such an aggregate, due to the particles' quasi-two-dimensional nature, they are on average all parallel to a mean planar orientation. The total three-dimensional assembly has to be understood as a combination of such aggregates, in which the possible mean planar orientations for the aggregates contain the direction of the electric field and are random otherwise. Consequently, the whole structure remains statistically unchanged by rotations around the axis of the electric field. Figure 1(c) shows a cut of the same aggregate through a plane parallel to its mean plane and indicated by the red dashed line in Fig. 1(b). Figures 1(b) and (c) correspond to two configurations met by the X-ray beam and that contribute equivalently to the scattering profiles recorded along the vertical direction.

In terms of intensity profiles along the principal axes of the two-dimensional SAXS images, the results can be understood as follows:

**Crossovers in the horizontal profiles:** Along the minor axis (horizontal direction), the two crossover $q$-values denote characteristic scales for the system: the larger one, $l_{\parallel 2}$, (see Fig. 4 (a), plot (1), and Table 1) is a typical particle extension/width, while the smaller one, $l_{\parallel 1}$, can be understood as a typical pore size between aggregated clay particles, along the direction of the external electric field (see Fig. 1(b) and (c)). Therefore, $l_{\parallel 1}$ is controlled by the particles' sizes and their relative orientations/ positioning, while $l_{\parallel 2}$ is less dependent on the particular arrangement of the particles with respect to each other (provided they are aligned with the electric field). We see that $l_{\parallel 1}$ is significantly different for the Na/Fe-fluorohectorite and the Ni-fluorohectorite samples. The parameter $l_{\parallel 1}$, which depends on the particle arrangement, can be influenced by the type of intercalated cation. From Table 1 we see that Ni-fluorohectorite also has $l_{\parallel 2}$ (average stack diameter) larger than for Na/Fe-fluorohectorite. Thus, it seems that this difference has an effect on the steric hindrances, and on the packing along the direction of the electric field, thereby increasing also the average pore size in the horizontal direction.

The typical relative length scales are indicated in Fig. 1(b); due to the particle polydispersity, they must be understood as average values that do not correspond to a true regular structure of the assembly. Such a regular structure would probably result in a correlation peak in the scattering data, which is not visible here.

**Crossovers in the vertical profiles:** Along the major axis (vertical direction, perpendicular to the electric field), due to the axial symmetry of the particle population around the direction of the electric field, some of the platelet-shaped particles present their width to the beam, and others present their top/bottom surfaces, so that both the particle- thickness and width should be visible on the scattering profiles. More precisely, if in Figure 1(b) and (c) one imagines an incident beam normal to the paper's plane, some aggregates are met by the beam as sketched in Fig. 1(b), others as sketched in Fig. 1(c):

- Intensity profiles resulting, along the vertical direction of the detector, from scattering by the latter aggregates (Fig. 1(c)), will exhibit two crossovers, one corresponding to the typical pore size $l\perp_1$, the other one to the typical particle size $l\perp_2$. Due to the statistical isotropy of the particles in their mean plane, $l\perp_2$ and $l_{||2}$ are actually identical (see Fig. 1(b-c)).
- Intensity profiles resulting, along the vertical direction of the detector, from scattering by the former aggregates (Fig. 1(b)), will exhibit two crossovers at the typical particle thickness $d$ and at a typical pore size in the direction perpendicular to the mean plane of an aggregate (vertical direction in Fig. 1(b)).

The profiles recorded along the vertical direction result from a mix of these two configurations and of all intermediate configurations in which the orientation of the aggregate's mean plane is in-between those of Fig. 1(b) and Fig. 1(c). Hence, at large scales we expect a crossover $q_{||2} = q\perp_2$. This is why we have used the value obtained from Fig. 4(a), where the crossover at small $q$ values is more marked. On the other hand, at small length scales, the relevant characteristic length scales are $d$, $l_{||1}$, $l\perp_1$, and the typical pore size along the normal to the mean plane of aggregate. We believe that these four length scales are close to each other, and are therefore not visible as distinct crossovers in the log-log intensity profiles. Thus, we have defined visually from Fig. 4(b) a single crossover at large $q$ values; its value is indeed close to that found for $q_{||1}$ in Fig. 4(a) (see Table 1).

**Signification of the exponents:** By changing $q$, we probe the system at different scales: for $q > q_1$ (in the two directions), we probe the properties of the individual particles; at smaller $q$- values, we probe the properties of the porous space between the particles, or of aggregates of particles. As summarized in section 2, the values of the exponents $D$ for the fitted power laws denote long range correlations in the inner surfaces (if $3 < D < 4$) or bulk arrangement (if $2 < D < 3$) of the porous medium. Furthermore, the value of exponents $D$ for the fitted power laws denote scattering from aggregated mass (if $1 < D < 3$) of particles if the probed length scales are larger than the particles size.

For $q > q_1$ (see $D_1$ values in Table 1, in the two directions), we expect the exponent to be controlled by the surface roughness of the particles for all the X-fluorohectorites samples, where X= $Na^+$, $Ni^{2+}$, and $Fe^{3+}$. This is what is observed here for Na/Ni-fluorohectorite samples, but not for Fe-fluorohectorite samples. The other exponents for Na/Ni-fluorohectorite samples are also greater than 3 in absolute value, which indicates a predominant surface scattering at all scales probed by the experiment for Na/Ni-fluorohectorite samples. Note that due to the axial symmetry discussed above, the exponent observed in Fig.4 (a), in the large $q$-range, is probably determined by the scattering properties of the lateral surfaces of the clay particles, while the exponent found in Fig. 4 (b), in the same range of $q$-values, results from a mixed scattering by the lateral- and the top/bottom surfaces of the platelet-shaped scatterer. This is why two different exponents are observed, although at those length scales smaller than the particle thickness the difference in meso-porous arrangement arising from the particles' anisotropy plays no role.

Fe- fluorohectorite samples behave differently from the other samples (Table 1): the exponent $D_{\parallel i}$ ($i=1, 2, 3$) and $D\perp_i$ ($i=1, 2$) are smaller than 3, indicating that, both along

minor- and major axis, the scattering is controlled by the mass aggregates of Fe-fluorohectorite particles, rather than by the roughness of their surface in bundle structure. Since the primary particles i.e., the stacked clay platelets, have the same chemical composition in all samples, this suggests that the nature of the intercalated cations changes the interactions between particles enough to influence the meso-structure of the aggregates.

The geometry studied here has similarities with that observed for dry-pressed samples, in which all particles have their flat surfaces parallel on average, in a nematic-like arrangement. Namely, we have inferred a geometric description in which the overall assembly possesses a statistical axial symmetry around the direction of the applied electric field, but consists of aggregates (like that sketched in Fig. 1(b)) with a nematic-like arrangement. Such nematic arrangements were previously studied using small angle neutron scattering (SANS) for Na-fluorohectorite samples, by Knudsen et al.[52]. In those samples, which were dehydrated under a uniaxial load of 0.12 MPa and 2.60MPa, and in which clay particles are expected to be much more densely packed than in the electro-rheological bundles addressed in the present paper, the authors inferred a correlation length for the porous medium in the direction of the applied load, and one along the average plane of the clay particles. These correlation lengths can be related to typical pore sizes in the two directions, and were observed to be 2 times larger along the particle planes than in the direction of the load, probably due to the particles' aspect ratio and the dense packing. In the present system, no typical pore size can be observed at length scales different from the typical pore size $l\perp_1$, i.e., in aggregates like that pictured in Fig. 1(b), the typical pore size normal to the aggregate's mean plane is similar to the pore size along that plane; this indicates a looser packing.

## IV. CONCLUSION

We have studied the porous medium in electrorheological bundles of fluorohectorite clay particles, using X-ray small angle scattering. The platelet-shaped scatterers have a lateral size in the micron range and a thickness 10−20 times smaller. Three types of samples were studied: Na-, Ni-, and Fe-fluorohectorite, corresponding to cations with three different valences, from +1 to +3. The two-dimensional SAXS patterns obtained from these bundled structures are clearly anisotropic, reflecting the preferential orientation of the particles in the field. They can be analyzed in order to obtain a geometrical characterization of the assembly of aggregated polarized particles. It confirms a geometric model for particle packing that we had previously inferred from WAXS experiments. Scattering profiles recorded along the azimuthal directions parallel to and perpendicular to the applied electric field exhibit several power law regimes, with crossover scales between them that correspond to typical length scales of the particle assembly. For samples of Na- and Ni-fluorohectorite, the exponents of the power laws are in the range usually attributed to surface scattering and mass scattering with a correlated roughness depending upon the nature of intercalated cations. For Fe-fluorohectorite, bulk scattering seems to be predominant, although this influence of the intercalated cations on the meso-structure of the bundles is not yet understood.

This study opens many new prospects. Firstly, one may combine SAXS with SALS studies in order to obtain power law exponents on an even wider $q$-range. Secondly, we plan to investigate the relationship between the local structure around the intercalated cation and the overall collective behavior of the clay particles, by studying some selected X-fluorohectorite samples (X being the intercalated cation) with EXAFS.

Thirdly, it will be of interest to relate the measured ellipse eccentricity to the orientation density probability function of the population of platelets, as done in a recent study[53] in which an identical geometry was addressed for magnetically-oriented Gibbsite particles.

## ACKNOWLEDGMENTS

The staff at the DUBBLE beamline at ESRF is gratefully acknowledged for support during the SAXS experiments. This work was supported by the Research Council of Norway (RCN) through a Strategical University Program and through the Nanomat Program: RCN projects 152426/431, 154059/420, 148865/432, and 138368/V30 and SUP154059/420.

**TABLE 1:** Calculated fit parameters: (i) average eccentricities of the two-dimensional scattering patterns, and (ii) power law scattering exponents $D_i$ (i = 1, 2, 3) and crossover $q_1$ and $q_2$ in the scattering profiles (Fig. 4)

| Sample | Na-fluorohectorite | Ni-fluorohectorite | Fe-fluorohectorite |
|---|---|---|---|
| E (V/mm) | 1000 | 1000 | 1000 |
| No. of images per curve | 7 | 6 | 9 |
| $e$ | 0.54 | 0.61 | 0.57 |
| Crossover $q_{\parallel 1}$ (nm$^{-1}$) | 0.086±0.003 | 0.055±0.002 | 0.087±0.005 |
| Crossover $l_{\parallel 1}$ (nm) | ~73±3 | ~114±5 | ~72±5 |
| Crossover $q_{\parallel 2}=q_{\perp 2}$ (nm$^{-1}$) | 0.028±0.002 | 0.025±0.001 | 0.028±0.002 |
| Crossover $l_{\parallel 2}=l_{\perp 2}$ (nm) | ~224±16 | ~252±10 | ~224±15 |
| Crossover $q_{\perp 1}$ (nm$^{-1}$) | 0.090±0.005 | 0.057±0.005 | 0.091±0.005 |
| Crossover $l_{\perp 1}$ (nm) | ~70±4 | ~110±5 | ~69±5 |
| $D_{\parallel 1}$ | 3.44±0.05 | 3.23±0.05 | 2.63±0.05 |
| $D_{\parallel 2}$ | 3.02±0.05 | 3.10±0.05 | 2.30±0.05 |
| $D_{\parallel 3}$ | 3.38±0.05 | 3.34±0.05 | 2.94±0.05 |
| $D_{\perp 1}$ | 3.38±0.05 | 3.21±0.05 | 2.60±0.05 |
| $D_{\perp 2}$ | 3.21±0.05 | 3.07±0.05 | 2.56±0.05 |

**FIG. 1: (a)** Stable electro-rheological structures of iron (Fe) - fluorohectorite clay particles in the host silicone oil and in the presence of a DC external electric field of a magnitude equal to 1 kV/mm. The chain bundles are along the direction of applied electric field. **(b)** A two-dimensional sketch of a nematic aggregate of clay particles inside an ER bundle; the direction of the external electric field is horizontal; the platelet-shaped clay particles are horizontal on average in this particular aggregate; many such aggregates, possessing various orientations around the direction of the electric field, form the overall particle bundle. **(c)** A horizontal cut of the same aggregate, for example through the plane denoted by the red dashed line in (b). The characteristic lengths $l_{\parallel 1}$ and $l_{\parallel 2}$ are also visible in this representation; due to the statistical isotropy of each particle in in its mean plane, the characteristic lengths $l_{\parallel 2}$ and $l_{\perp 2}$ are identical.

**FIG. 2**: Experimental setup for recording SAXS data from the electro-rheological chain bundles of clay particles. Y is the sample to detector distance (camera length).

**FIG. 3:** Contour plot of a two-dimensional anisotropic scattering pattern obtained from bundles of iron (Fe)-fluorohectorite clay particles initially dispersed in silicone oil (at 3.0% w/w), in the presence of a DC external electric field of magnitude equal to 1 kV/mm (see Fig. 1). Each of the elliptically-shaped black areas corresponds to an iso-intensity region, i.e. to a given detector count; the logarithms of the detector count values for these iso-intensity regions are equally-spaced. The raw image data, in which grey shadings are proportional to the detector count, is shown as an inset in the top righ part of the image.

**FIG. 4:** Log-Log plots of the intensity of the scattering profiles recorded as a function of the scattering vector $q$. The curves have been shifted vertically for better visualization. Different slopes are observed for (1) sodium (Na)-fluorohectorite, (2) nickel (Ni)-fluorohectorite, and (3) iron (Fe)-fluorohectorite, with different crossover between them (see Table 1). **(a):** along the minor axis (horizontal direction, i.e., direction of the applied electric field) and **(b):** along the major axis (vertical axis).

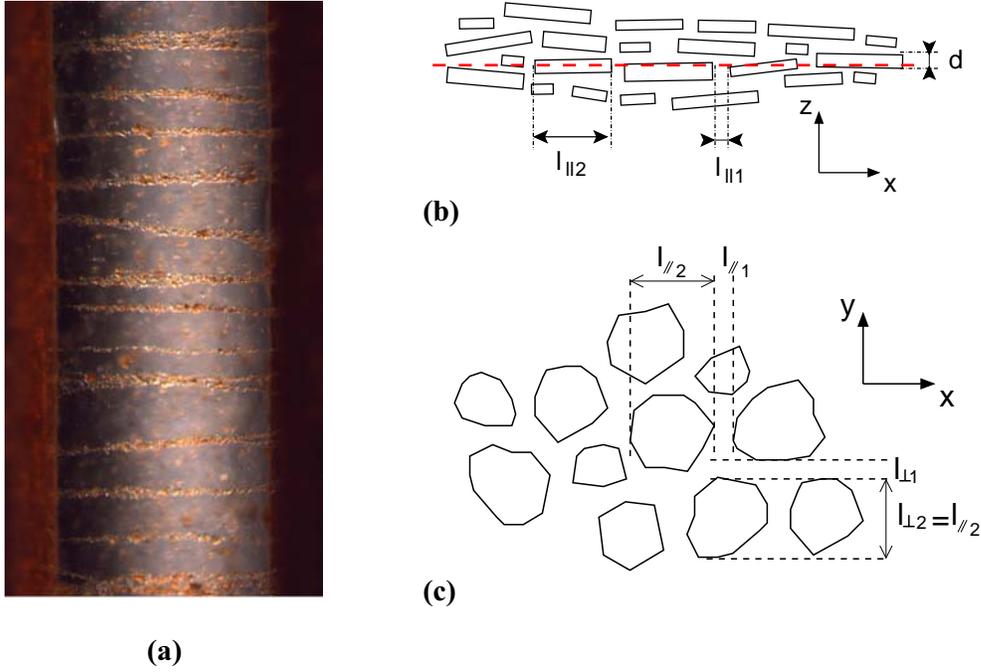

**FIG. 1: (a)** Stable electro-rheological structures of iron (Fe) - fluorohectorite clay particles in the host silicone oil and in the presence of a DC external electric field of a magnitude equal to 1 kV/mm. The chain bundles are along the direction of applied electric field. **(b)** A two-dimensional sketch of a nematic aggregate of clay particles inside an ER bundle; the direction of the external electric field is horizontal; the platelet-shaped clay particles are horizontal on average in this particular aggregate; many such aggregates, possessing various orientations around the direction of the electric field, form the overall particle bundle. **(c)** A horizontal cut of the same aggregate, for example through the plane denoted by the red dashed line in (b). The characteristic lengths $l_{\|1}$ and $l_{\|2}$ are also visible in this representation; due to the statistical isotropy of each particle in in its mean plane, the characteristic lengths $l_{\|2}$ and $l_{\perp 2}$ are identical.

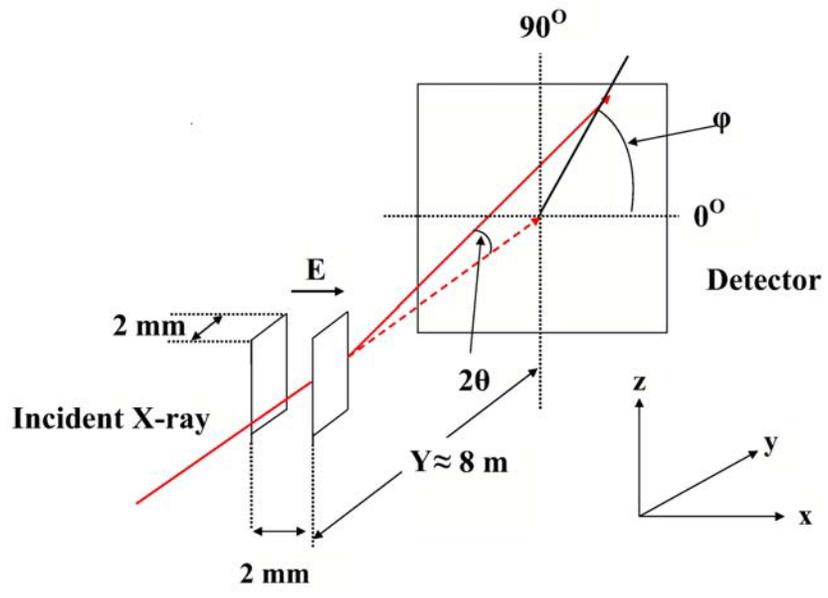

**FIG. 2**: Experimental setup for recording SAXS data from the electro-rheological chain bundles of clay particles. Y is the sample to detector distance (camera length).

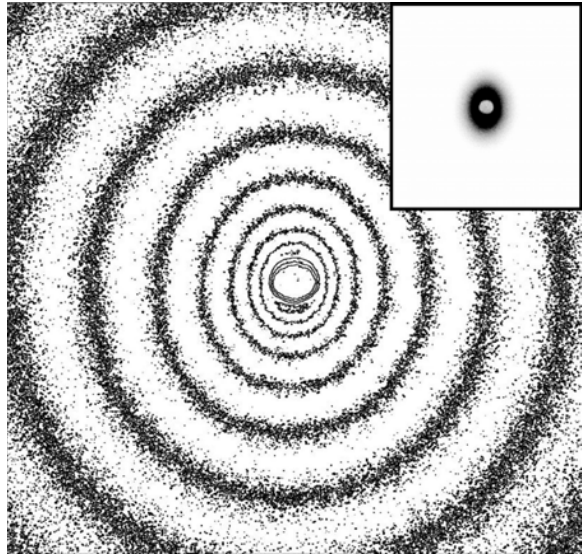

**FIG. 3:** Contour plot of a two-dimensional anisotropic scattering pattern obtained from bundles of iron (Fe)-fluorohectorite clay particles initially dispersed in silicone oil (at 3.0% w/w), in the presence of a DC external electric field of magnitude equal to 1 kV/mm (see Fig. 1). Each of the elliptically-shaped black areas corresponds to an iso-intensity region, i.e. to a given detector count; the logarithms of the detector count values for these iso-intensity regions are equally-spaced. The raw image data, in which grey shadings are proportional to the detector count, is shown as an inset in the top righ part of the image.

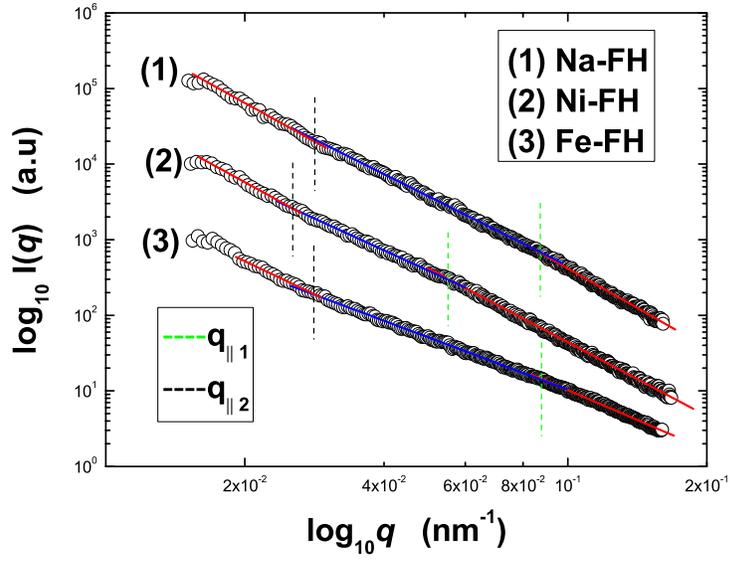

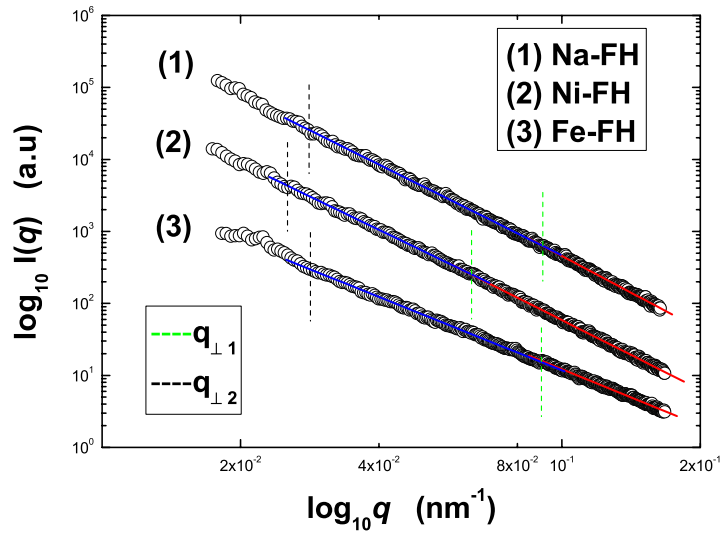

**FIG. 4:** Log-Log plots of the intensity of the scattering profiles recorded as a function of the scattering vector $q$. The curves have been shifted vertically for better visualization. Different slopes are observed for (1) sodium (Na)-fluorohectorite, (2) nickel (Ni)-fluorohectorite, and (3) iron (Fe)-fluorohectorite, with different crossover between them (see Table 1). **(a):** along the minor axis (horizontal direction, i.e., direction of the applied electric field) and **(b):** along the major axis (vertical axis).